\pdfoutput=1 
\documentclass[reprint, amsmath,amssymb,aps,pra]{revtex4-1}
\usepackage{graphicx}
\usepackage{subfigure}
\usepackage{morefloats}
\usepackage{bm}
 \usepackage{comment}
\normalsize 
\large
\begin{document}
\title{Quantum Simulation of Dzyaloshinsky-Moriya Interaction}
\author{V. S. Manu }
\author{Anil Kumar} 
\affiliation{Centre for Quantum Information and Quantum Computing, Department of Physics and NMR Research Centre,
Indian Institute of Science,  Bangalore-560012}
\begin{abstract}
Quantum simulation of a Hamiltonian H requires unitary operator decomposition (UOD) 
of its evolution operator, ($U=exp(-i H t)$) in terms of experimentally preferable 
unitaries. Here, using Genetic Algorithm optimization, we numerically evaluate the 
most generic UOD for the Hamiltonian, DM interaction in the presence of Heisenberg 
XY interaction, $H_{DH}$. Using these decompositions, we studied the entanglement dynamics of 
Bell state in the Hamiltonian $H_{DH}$ and verified the entanglement preservation procedure by Hou et al. [\textit{Annals of Physics, \textbf{327} 292 (2012)}].
\end{abstract}
\pacs{}
\maketitle
\section{Introduction} \label{sec:intro}

Algorithms with exponential speedups over classical counterparts \cite{shorsalgorithm,groversalgorithm}, 
simulation of quantum systems \cite{sethlloyd, du_hydrogen,lidar_thermalrate} and testing basic 
principles of quantum mechanics \cite{jharana_nohiding,mahesh_legget} makes 
quantum Information processing (QIP) and quantum computation  intensively investigated fields of physics over last decade. 
The idea of simulating quantum systems in a quantum computer was proposed 
by Feynman \cite{feynman_qs} in 1982 and is one of the most important practical application of the 
quantum computer. Quantum simulation has the potential to revolutionize the physics and chemistry and draws attention 
recently by solving problems like -- molecular Hydrogen simulation \cite{du_hydrogen}, 
calculations of thermal rate constants of chemical reactions \cite{lidar_thermalrate} and quantum chemical dynamics 
\cite{kassal_chemicaldynamics}.

Dzyaloshinsky-Moriya (DM) interaction is an anisotropic antisymmetric exchange interaction arising from spin-orbit 
coupling \cite{dzyaloshinsky, moriya}.
It was proposed by Dzyaloshinski to explain the weak ferromagnetism of antiferromagnetic crystals 
($\alpha$-$Fe_2O_3$, $MnCO_3$)\cite{dzyaloshinsky}.
DM interaction is crucial in the description of many antiferromagnetic systems \cite{dender,kohgi,greven} and is 
important in the entanglement properties of the system. Here we present 
a generic unitary operator decomposition which will help to simulate the Hamiltonian -- DM interaction in the presence of Heisenberg 
XY interaction -- in a two qubit system with almost any basic interaction between them.

Long-lasting coherence and high fidelity controls in nuclear magnetic resonance (NMR) are ideal for quantum 
information processing. Experimental implementation quantum algorithms  
(Deutsch-Jozsa algorithm, Grover's search algorithm and Shor's algorithm of factorization), testing basic 
principles of quantum mechanics (nohiding theorem \cite{jharana_nohiding} and Leggett-Garg inequality 
\cite{mahesh_legget}) and quantum simulation 
(hydrogen molecule \cite{du_hydrogen} and system with competing two and three Body interactions 
\cite{suter_2and3bodyinteractions}) were performed in Liquid state NMR (LSNMR).

Genetic algorithms (GA) are stochastic global search method based on the mechanics of natural biological 
evolution \cite{whitely_ga}. It was first proposed by John Holland in 1975 \cite{holland_ga}. GA operates on 
a population of solutions of a specific problem by encoding the solutions to a simple chromosome-like 
data structure, and applies recombination operators. GAs are attractive in engineering design and applications because 
they are easy to use and are likely to find the globally best design or solution, which is superior to 
other design or solution \cite{rasheed_ga}. Here we used Genetic algorithm optimization for solving 
UOD for generic DM Hamiltonian with Heisenberg-XY interaction. 

Section \ref{sec:theory} deals with theoretical discussion of DM Hamiltonian simulation followed by experimental 
implementation in Section \ref{sec:expt}.

\section{Theory} \label{sec:theory}

DM interaction in the presence of Heisenberg XY interaction $H(J,D)$ is, 
\begin{equation} \label{eqn:dmh}
H(J,D)=J(\sigma_{1x}\sigma_{2x}+\sigma_{1y}\sigma_{2y})+D(\sigma_{1x}\sigma_{2y}-\sigma_{1y}\sigma_{2x}),
\end{equation}
where J and D respectively represents the strength of Heisenberg and DM interactions.

Experimental simulation of $H(J,D)$ (Eqn. \ref{eqn:dmh}) in a quantum system (with Hamiltonian $H_{sys}$)  requires 
UOD of evolution operator $U(J,D,t)$,
\begin{equation} \label{eqn:dmu}
U(D,J,t)=exp(-i H(J,D) \times t),
\end{equation}
in terms  of Single Qubit Rotations 
$R^n(\theta,\phi)$ ($\theta$ angle rotation along $\phi$ axis on $n^{th}$ spin),
\begin{equation} \label{eqn:usqr}
R^n(\theta,\phi) = exp(-i \theta/2 \times [Cos\phi ~\sigma_{nx} +Sin\phi ~\sigma_{ny}]),
\end{equation}
and evolution under system Hamiltonian $U_{sys}$  (Eqn. \ref{eqn:sysu}),
\begin{equation} \label{eqn:sysu}
U_{sys}(t)=exp(-i H_{sys} \times t).
\end{equation}

Without losing generality, Eqn. \ref{eqn:dmu} can be written  as,
\begin{equation} \label{eqn:dmu1}
\begin{split}
U(\gamma,\tau)=exp(-i [(\sigma_{1x}\sigma_{2x}+\sigma_{1y}\sigma_{2y})+ \\ 
\gamma (\sigma_{1x}\sigma_{2y}-\sigma_{1y}\sigma_{2x})]~ \tau),
\end{split}
\end{equation}
where $\gamma$ represents the relative ratio of interaction strengths ($\gamma = D/J$) and $\tau=J\times t$.

Eqn. \ref{eqn:dmu1} forms the complete unitary operator for the Hamiltonian (Eqn. \ref{eqn:dmh}) with  
$\gamma$ and $\tau$ varies from 0 to $\infty$. We performed UOD for Eqn. \ref{eqn:dmu1} using 
Genetic algorithm optimization \cite{vsmanu_ga}.
In an operator optimization (as shown in \cite{vsmanu_ga}), optimization is performed for a constant unitary 
matrix-- corresponds to a single fidelity point. 
Here optimization has to be performed for a two dimensional fidelity profile generated by $\gamma$ and $\tau$. We name it as Fidelity Profile 
Optimization (FPO). FPO for the present case is explained in following steps.

In the first step, we performed Fidelity Profile Optimization with following assumptions -- (a).the range of 
$\tau$ is from 0 to 15, (b). the range of $\gamma$ is from 0 to 1 and (c). the system Hamiltonian ($H_{sys}$) is 
given by Eqn. \ref{eqn:zzh}. 
\begin{equation} \label{eqn:zzh}
H_{sys}=H_{zz}=J_{zz} (\sigma_{1z}\sigma_{2z}).
\end{equation}
where $J_{zz}$ is the strength of zz-interaction.

The optimization procedure using Genetic algorithm is explained in the Supporting information. 

\begin{figure}
\includegraphics[width=0.35\textwidth, height=0.27\textwidth]{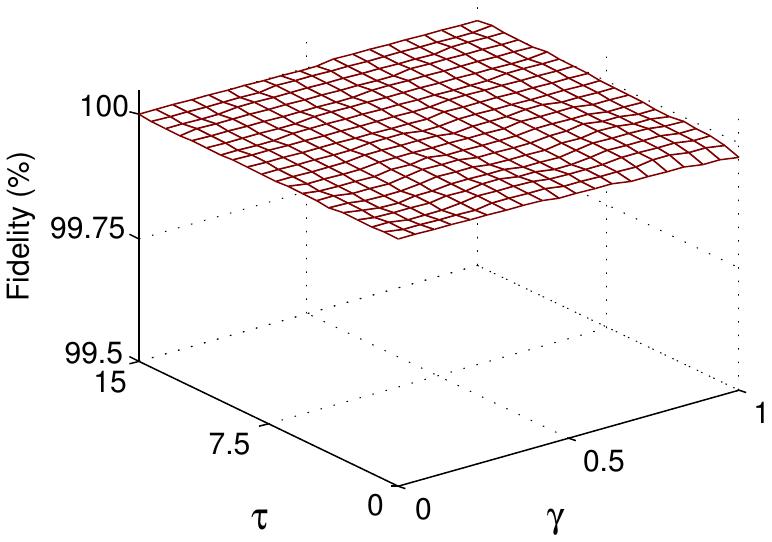}
\caption{Fidelity profile of UOD given in Eqn. \ref{eqn:uh1}.}  \label{fig:fp1}
\end{figure}

The optimized UOD (Eqn. \ref{eqn:uh1}) has seven SQRs and two system Hamiltonian evolutions. 
\begin{equation} \label{eqn:uh1}
\begin{split}
U(\gamma,\tau) = R^1(\tfrac{\pi}{2},-\tfrac{\pi}{2})  R^1(\tfrac{\pi}{2},\theta_2) R^2(\pi,\pi)  	
U_{zz}(\tfrac{\pi}{4})\\  R^1(\theta_1,\theta_2+\tfrac{\pi}{2})R^2(\pi-\theta_1,0)			
U_{zz}(\tfrac{\pi}{4})\\  R^1(\tfrac{\pi}{2},\theta_2+\pi) R^2(\tfrac{\pi}{2},\tfrac{\pi}{2}), 
\end{split}
\end{equation}
where  $\theta_1$ and $\theta_2$ (Eqn. \ref{eqn:th1}) impart  $\gamma$ and $\tau$ dependence to UOD 
and $U_{zz}$ is given by Eqn. \ref{eqn:sysu} with $J_{zz} \times t =\pi/4$.
\begin{equation} \label{eqn:th1}
\begin{split}
\theta_1=[0.8423-0.3455~Cos(1.117~\gamma)+ \\0.01806~Sin(1.117 \gamma)]~\tau \\
\theta_2=1.345~exp(-0.8731 \gamma)+1.796.
\end{split}
\end{equation}
The fidelity \cite{vsmanu_ga} profile of UOD is 
shown in Fig. \ref{fig:fp1}. The minimum point in fidelity profile is greater than 99.99 \%. 
It should be noted that, the total length of UOD (Eqn. \ref{eqn:uh1}) 
is invariant under $\gamma$ and $\tau$ (with the assumption -- all the SQRs are instantaneous).

For generalizing the assumption on $\tau$, we solved Eqn. \ref{eqn:pf} numerically and find the 
period P($\gamma$) of U($\gamma$,$\tau$) (Eqn. \ref{eqn:p}). 
\begin{equation} \label{eqn:pf}
 H(\gamma,\tau+n \times P(\gamma))= H(\gamma,\tau).
\end{equation}
\begin{equation} \label{eqn:p}
 P(\gamma)=3.008~\gamma^3-6.627~\gamma^2-0.1498~\gamma+12.59.
\end{equation}
Eqn. \ref{eqn:p} has a maximum value of 12.59 at $\gamma$=0. Since the maximum value of period is 
less than 15 (FOP performed till $\tau$=15), UOD (Eqn. \ref{eqn:uh1}) can be used for any value of $\tau$.
Same argument can be used for extending the range of $\tau$ to $-\infty$.

In order to incorporate the range of $\gamma$ from 0 to $\infty$, we performed  FPO for the 
operator Eqn. \ref{eqn:dmu2}),
\begin{equation}\label{eqn:dmu2}
\begin{split} 
U'(\gamma',\tau')=exp(-i [\gamma' (\sigma_{1x}\sigma_{2x}+\sigma_{1y}\sigma_{2y})+ \\ 
(\sigma_{1x}\sigma_{2y}-\sigma_{1y}\sigma_{2x})]~ \tau'),
\end{split}
\end{equation}
and the optimized unitary decomposition are,
\begin{equation} \label{eqn:uh2}
\begin{split}
U'(\gamma',\tau') = R^1(\tfrac{\pi}{2},\tfrac{\pi}{2}) R^2(\tfrac{\pi}{2},\theta_3)  	
U_{zz}(\tfrac{\pi}{4}) R^1(\theta_2+\theta_3,0)\\ R^2(\theta_1,\theta_4)			
U_{zz}(\tfrac{\pi}{4}) R^1(\tfrac{\pi}{2},\tfrac{\pi}{2}) R^2(\tfrac{\pi}{2},\theta_3),
\end{split}
\end{equation}
where  $\theta_1 \cdots \theta_4$ (Eqn. \ref{eqn:th2}) impart  $\gamma$ and $\tau$ dependence to UOD.

\begin{small}\begin{align} \label{eqn:th2}
\begin{split}
&\theta=[ 0.09812~ exp(-2.42 \gamma)+0.4023 ~exp(0.5524 \gamma)] \tau,\\
&\theta_1=-\theta+3.142, \\
&\theta_2=\theta-[1.242 ~exp(-0.9617 \gamma)+ 0.3546~exp(-0.1145 \gamma)], \\
&\theta_3=1.259~exp(-0.957 \gamma)+3.479~exp(-0.0087 \gamma), \\
&\theta_4=1.256~exp(-0.959 \gamma)+ 1.912~exp(-0.0166 \gamma),
\end{split}
\end{align}
\end{small}
where $\gamma'$ varies from 0 to 1 and $\tau$ from 0 to 15.

Eqn. \ref{eqn:dmu2} satisfy the same periodicity relation as shown in Eqn. \ref{eqn:p} and hence 
can use the same reasoning for extending $\tau$ range from 0 to $+\infty$.

For $\gamma > 1$, Eqn. \ref{eqn:dmu1} can be written as,
\begin{equation} \label{eqn:dmut}
\begin{split}
U(\gamma'',\tau'')=exp(-i [\gamma''(\sigma_{1x}\sigma_{2x}+\sigma_{1y}\sigma_{2y})+ \\ 
 (\sigma_{1x}\sigma_{2y}-\sigma_{1y}\sigma_{2x})]~  \tau''),
\end{split}
\end{equation}
where $\gamma''=1/\gamma$ and $\tau''=\gamma \times \tau$.

Eqn. \ref{eqn:dmu2} and Eqn. \ref{eqn:dmut} are equivalent and hence UOD for Eqn. \ref{eqn:dmu1} can be shown as,
\begin{equation}\label{eqn:uod}
U(\gamma, \tau) =   \left\{ 
  \begin{array}{l l}
    \text{Eqn. \ref{eqn:dmu1}} & \quad \text{if $\gamma \leqslant $  1}\\
    \text{Eqn. \ref{eqn:dmu2}} & \quad \text{if $\gamma >$  1}\\
  \end{array} \right.
\end{equation}

The UOD optimization given  Eqn. \ref{eqn:uh1} is based on $H_{sys}=H_{zz}$ (Eqn. \ref{eqn:zzh}). 
It can be generalized to almost any interaction by \textit{term isolation} procedure by Bremner et al. \cite{bremner,hill}. 

As an example consider the case,
\begin{equation} \label{eqn:hsystst}
H_{sys}= J(\sigma_x\sigma_x+\sigma_y\sigma_y+\sigma_z\sigma_z).
\end{equation}

The $H_{zz}$ terms can be isolated from Eqn. \ref{eqn:hsystst} and is shown in Eqn. \ref{eqn:hsyststiso}.

\begin{equation} \label{eqn:hsyststiso}
\begin{split}
exp(-iJ_{zz}\sigma_z\sigma_z t)=R^1(\pi,z)exp(-iH_{sys}t)R^1(\pi,z)\\exp(-iH_{sys}t),
\end{split}
\end{equation}
where $R^1(\pi,z) $ represents a $\pi$-SQR on spin 1 along Z axis.

Combinig all the steps above forms most generic UOD of the Hamiltonian -- DM interaction in the presence of Heisenberg XY interaction.

\section{Experimental Quantum Simulation} \label{sec:expt}
\begin{figure}
\includegraphics[width=0.14\textwidth, height=0.12\textwidth]{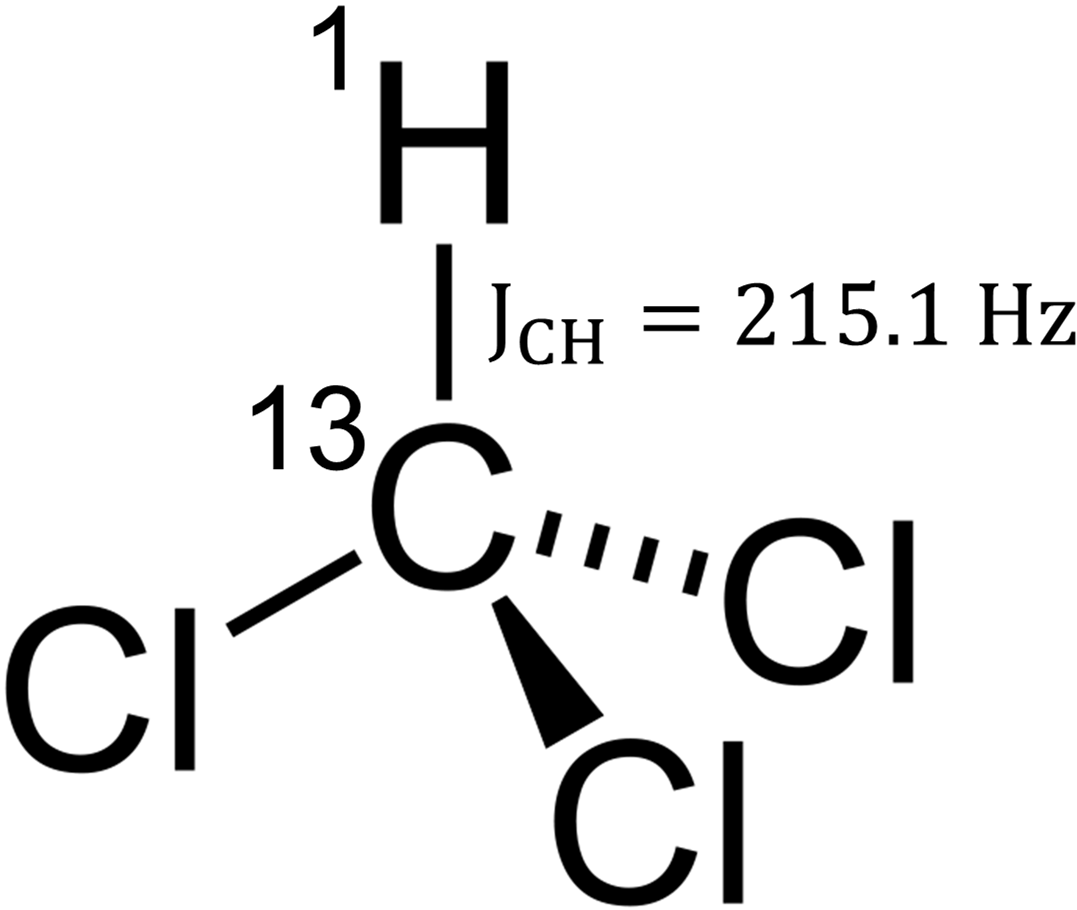}
\caption{$^{13}C$ labeled Chloroform used for quantum simulation. $^{13}C$ and $^1H$ act as qubits with zz 
interaction ($J_{CH}$=215.1 Hz) between them.}  \label{fig:chcl3}
\end{figure}

We performed Quantum simulation experiments in a two qubit NMR system $^{13}CHCl_3$ (dissolved in Acetone-D6) 
(Fig. \ref{fig:chcl3}) with $^{13}C$ and $^1H$ spins act as two qubit system with scalar coupling (zz interaction-- Eqn. \ref{eqn:zzh}) 
between them. The system Hamiltonian is zz interaction (Eqn. \ref{eqn:zzh}). We performed all the experiments in Bruker AV-500 
spectrometer.

Quantum computation experiments in NMR starts with (i). preparation of pseudo pure states 
\cite{cory_pps, gershenfeld_pps, Mahesh_sallt}, (ii). processing the state by evolving under different average 
Hamiltonians \cite{ernstbook} and (iii). read-out by quantum state tomography \cite{avik_adiabatic}. 
Here we studied the entanglement dynamics (quantified by concurrence \cite{coffman_concurrence}) of a Bell state 
(Eqn. \ref{eqn:bell2}) in the Hamiltonian given in  Eqn. \ref{eqn:dmh}.

\begin{equation} \label{eqn:bell2}
 |\phi\rangle_-=\frac{1}{\sqrt{2}}(|01\rangle-|10\rangle)
\end{equation}

Using the unitary operator decompositions shown in Eqn. \ref{eqn:uod}, we have simulated the Hamiltonian $H(\gamma, \tau)$ 
for $\gamma$=\{0.33, 0.66, 0.99\} and studied the entanglement dynamics of the singlet Bell state (Eqn. \ref{eqn:bell2}) under these Hamiltonians.

In experimental implementation $R^n(\theta,\phi)$ is implemented by hard pulse \cite{spindynamics} with suitable length 
(determined by $\theta$) along the axis $\phi$ on $^1H$ or $^{13}C$ spin and $U_{zz}(\theta')$ is implemented by system 
Hamiltonian evolution for a time determined by $\theta'$ \cite{ernstbook}. The experimental simulation results 
(Fig. \ref{fig:ed})  shows a good agreement with the theoretical simulation. Average experimental deviation ($AED$) 
in concurrence -- calculated 
using the formula Eqn. \ref{eqn:cd} -- is 3.83\%, which is mainly due to decoherence effects, static and rf inhomogeneities.

\begin{equation} \label{eqn:cd}
 AED= \displaystyle \sum_{i=1}^{n} \frac{|C_{es}(i)-C_{ts}(i)|}{C_{ts}(i)}
\end{equation}

where $C_{es}(i)$ and $C_{ts}(i)$ are the concurrence in experimental and theoretical simulations and $n$ is the 
number of experimental points (here we performed simulation for $n=16$).
\subsubsection{Entanglement Preservation}

Hou et al. \cite{hou2012preservation} demonstrated a mechanism for entanglement preservation of a quantum state in a Hamiltonian of the 
type given in Eqn. \ref{eqn:dmu}. 

Preservation of initial entanglement of a  quantum state is performed by free evolution interrupted with a certain operator $O$, which makes the 
state to go back to its initial state. The operator sequence for preservation is given in Eqn. \ref{eqn:epp}.

\begin{equation} \label{eqn:epp}
 OU{(\gamma, \tau)}OU{(\gamma, \tau)}\equiv I,
\end{equation}
where $O=I_1\otimes \sigma_{2z}$.

\begin{figure}[t]
 \subfigure[~]{\includegraphics[width=0.35\textwidth, height=0.25\textwidth]{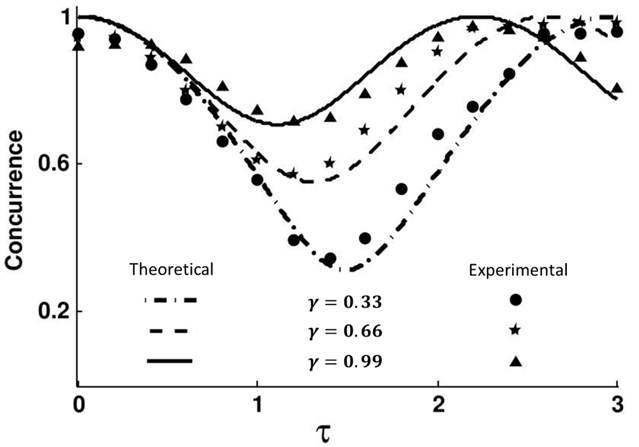}  \label{fig:ed}} \\
 \subfigure[~]{\includegraphics[width=0.35\textwidth, height=0.25\textwidth]{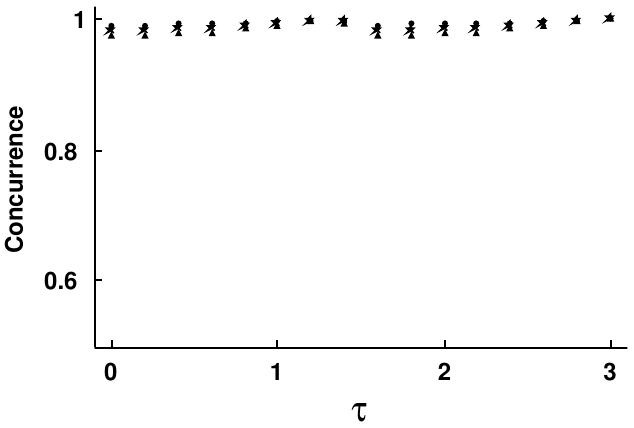} \label{fig:ep}}
\caption{(a).Entanglement (concurrence) dynamics of $^{13}C-H$ system under the Hamiltonian Eqn. \ref{eqn:dmh}, 
(b). Entanglement preservation experiment using  Eqn. \ref{eqn:epp}. Starting from singlet state, the concurrence 
sustains at 1 with the preservation procedure.}
\end{figure}

We performed entanglement preservation experiment for singlet state (Eqn. \ref{eqn:bell2}) in $H(\gamma,\tau)$ with $\gamma$=\{0.33, 0.66, 0.99\}.
The experimental results (Fig. \ref{fig:ep}) shows excellent entanglement preservation and good agreement with the theoretical simulation. 
The experimental deviation of concurrence (Eqn. \ref{eqn:cd}) is less than 2\%.

\section*{Conclusion}
We have performed Fidelity Profile Optimization for the Hamiltonian -- DM interaction in the presence 
of Heisenberg XY interaction. The optimized UOD can be used for all  relative strengths ($\gamma$) of the  
interactions and length is invariant under $\gamma$ or evolution time. Using these decompositions, 
we have experimentally verified the entanglement preservation mechanism 
suggested by Hou et al.

\section*{Acknowledgments}

We thank Prof. Apoorva Patel for discussions and suggestions,
and the NMR Research Centre for use of the AV-500 NMR spectrometer.
V.S.M. thanks UGC-India for support through a fellowship.

\bibliographystyle{unsrtnat}
\bibliography{manu_DM}
\end{document}